\documentclass[pra,twocolumn,english,reprint, longbibliography, superscriptaddress, breaklinks=true, showkeys, showpacs=false, nofootinbib]{revtex4-2}
\usepackage[T1]{fontenc}
\usepackage[utf8]{inputenc}
\usepackage{color}
\usepackage{babel}
\usepackage{amsmath}
\usepackage{amssymb}
\usepackage{subfigure}
\usepackage{graphicx}
\usepackage{physics} 
\usepackage{dsfont}
\usepackage{hyperref}
\usepackage{comment}
\usepackage{tabularx} 
\usepackage{soul}
\newcolumntype{Y}{>{\centering\arraybackslash}X}

\usepackage{listings}
\usepackage{xcolor} 
\usepackage{ragged2e}
\usepackage{bbding}
\newcommand{\greencheck}{{\color{green}\CheckmarkBold}}
\newcommand{\redcross}{{\color{red}\XSolidBold}}

\begin{document}

\author{Lucas Friedrich\href{https://orcid.org/0000-0002-3488-8808}{\includegraphics[scale=0.05]{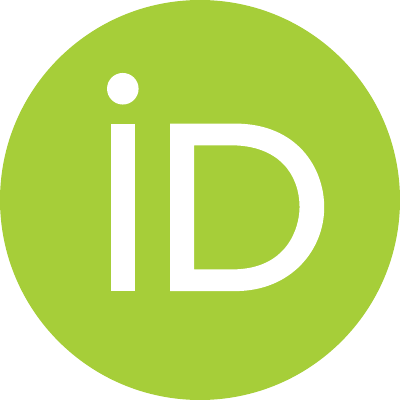}}}
\email{lucas.friedrich@acad.ufsm.br}
\affiliation{Physics Department, 
Federal University of Santa Maria, 97105-900,
Santa Maria, RS, Brazil}

\author{Marcos L. W. Basso\href{https://orcid.org/0000-0001-5456-7772}{\includegraphics[scale=0.05]{orcidid.pdf}}}
\email{marcoslwbasso@hotmail.com}
\affiliation{Center for Natural and Human Sciences, Federal University of ABC, Santo Andr\'e, SP, 09210-580, Brazil}
\affiliation{Department of Applied Mathematics, State University of Campinas, 13083-859 Campinas, S\~ao Paulo, Brazil}

\author{Alberto B. P. Junior \href{https://orcid.org/0000-0003-1942-245X}{\includegraphics[scale=0.05]{orcidid.pdf}}}
\email{alberto.palhares.112@ufrn.edu.br}
\affiliation{Physics Department, Federal University of Rio Grande do Norte, Natal, 59072-970, Rio Grande do Norte, Brazil}
\affiliation{International Institute of Physics, Federal University of Rio Grande do Norte, 59078-970, Natal, Brazil}

\author{Joab M. Varela
\href{https://orcid.org/0000-0001-9182-6104}
{\includegraphics[scale=0.05]{orcidid.pdf}}}
\email{joabapc16@gmail.com}
\affiliation{Physics Department, Federal University of Rio Grande do Norte, Natal, 59072-970, Rio Grande do Norte, Brazil}
\affiliation{International Institute of Physics, Federal University of Rio Grande do Norte, 59078-970, Natal, Brazil}

\author{Leandro Morais
\href{https://orcid.org/0009-0005-5773-1116}
{\includegraphics[scale=0.05]{orcidid.pdf}}}
\email{leandro.silva@ifsc.usp.br}
\affiliation{International Institute of Physics, Federal University of Rio Grande do Norte, 59078-970, Natal, Brazil}
\affiliation{Instituto de Física de São Carlos, Universidade de São Paulo, CP 369, 13560-970 São Carlos, SP, Brazil}

\author{Rafael Chaves\href{https://orcid.org/0000-0001-8493-4019}{\includegraphics[scale=0.05]{orcidid.pdf}}}
\email{rafael.chaves@ufrn.br}
\affiliation{International Institute of Physics, Federal University of Rio Grande do Norte, 59078-970, Natal, Brazil}
\affiliation{School of Science and Technology, Federal University of Rio Grande do Norte, Natal, Brazil}

\author{Jonas Maziero\href{https://orcid.org/0000-0002-2872-986X}{\includegraphics[scale=0.05]{orcidid.pdf}}}
\email{jonas.maziero@ufsm.br}
\affiliation{Physics Department, 
Federal University of Santa Maria, 97105-900,
Santa Maria, RS, Brazil}

\selectlanguage{english}%

\title{A variational quantum algorithm for entanglement quantification}

\begin{abstract}
Quantum entanglement is a foundational resource in quantum information science, underpinning applications across physics. However, detecting and quantifying entanglement remains a significant challenge. In this article, we introduce a variational quantum algorithm inspired by Uhlmann’s theorem to quantify the Bures entanglement of general quantum states, a method that naturally extends to other quantum resources, including genuine multipartite entanglement, quantum discord, quantum coherence, and total correlations, while also enabling reconstruction of the closest free states. The algorithm requires a polynomial number of ancillary qubits and circuit depth relative to the system size, dimensionality, and free state cardinality, making it scalable for practical implementations. Thus, it provides a versatile framework for quantifying quantum resources, demonstrated here through several applications.
\end{abstract}


\date{\today}

\maketitle

\section{Introduction}
\label{sec:intro}

Quantum entanglement is a fundamental feature of quantum mechanics, first recognized in the early 20th century through the seminal works of Schrödinger~\cite{Schrodinger} and Einstein, Podolsky, and Rosen~\cite{EPR}. Since then, entanglement has been the subject of extensive theoretical and experimental investigation, enabling key advancements in quantum information processing~\cite{Bennett,Wiesner,Ekert,Shor} and related areas \cite{Amico,Erhard}. The reliable detection and precise quantification of entanglement are crucial for the practical implementation of these technologies. While entanglement witnesses~\cite{Horodecki,Guhne} provide a relatively straightforward method for detecting entanglement, developing a more efficient, accurate, and broadly applicable approach to quantifying entanglement, particularly in mixed multipartite states, remains a major challenge in quantum information science.

The difficulty in quantifying entanglement arises from the intricate nature of mixed-state entanglement~\cite{Bengtsson}. Unlike pure states, where measures such as the von Neumann entropy of subsystems provide a straightforward characterization, mixed-state entanglement lacks a universally accepted measure that is both computable and operationally meaningful, with several measures being introduced from convex roof and distance-based constructions~\cite{Plenio, Guhne, Horodecki}. Among these measures, the Bures measure of entanglement~\cite{Vedral97, Streltsov, Bromley}, defined from the Bures metric, provides a distance-based measure to quantify entanglement. However, as with the other entanglement measures, calculating the Bures entanglement involves solving an optimization problem over the set of separable mixed states, a task that is computationally prohibitive for large systems.


On the other hand, variational quantum algorithms (VQAs)\cite{CEREZO_vqa} have established themselves as versatile and promising tools for tackling complex problems in quantum information science, particularly in the context of Noisy Intermediate-Scale Quantum (NISQ) devices \cite{Bharti}. By leveraging a hybrid approach that combines parameterized quantum circuits with classical optimizers, these algorithms exhibit resilience to certain types of noise and adaptability to a wide range of problems. This flexibility has enabled successful applications in areas such as quantum simulation \cite{Bauer}, solving linear systems \cite{Harrow}, and quantum machine learning \cite{Hur_2022,Perez_Salinas,Li_2023,Fan_2023,Sagingalieva_2023}. Notably, VQAs have also been explored as viable approaches for entanglement quantification \cite{Chen}, with proposals ranging from variational estimation of logarithmic negativity \cite{Wang} to algorithms for quantum steering detection \cite{Philip} and variational determination of the geometric measure of entanglement~\cite{Moller22} while avoiding barren plateaus \cite{Zambrano}. Within this context, our goal in this paper is to employ VQA's as an operationally meaningful, accurate and feasible manner to compute the Bures entanglement of arbitrary quantum states.

The key insight in our approach is the use of Uhlmann's theorem~\cite{uhlmann, jozsa}, which is reviewed in Appendix \ref{sec:uhlmann} and states that the fidelity between two mixed states can be obtained by maximizing the overlap between their purifications in an extended Hilbert space.
In this way, Uhlmann's theorem provides a natural framework for computing the Bures measure of entanglement by finding the optimal purification via variational quantum circuits. Our method leverages the swap test~\cite{Barenco, Buhrman, li}, which is reviewed in Appendix \ref{sec:swap} and is a fundamental quantum subroutine for estimating inner products between quantum states, to evaluate overlaps between purifications, while avoiding quantum state tomography. By incorporating the swap test into the VQA framework, we introduce a scalable, polynomially bounded method for quantifying entanglement in bipartite and multipartite systems.

Furthermore, as detailed in Appendix \ref{Sec:resource}, our approach extends naturally beyond entanglement to other quantum resource theories, including genuine multipartite entanglement in Appendix \ref{subsec:gmeq}, quantum discord in Appendix \ref{subsec:qdq}, quantum coherence in Appendix \ref{subsec:qcq}, and total correlations in Appendix \ref{subsec:tcq}. In addition to quantification, our method enables reconstruction of the closest free states, such as separable states for entanglement, providing deeper insights into the structure of quantum correlations. The scalability and adaptability of our algorithm make it a promising candidate for practical implementations on near-term quantum devices, paving the way for new advances in entanglement theory and quantum information processing.

The sequence of this article is organized as follows. In Sec. \ref{sec:evqa}, we introduce the entanglement variational quantum algorithm. The results obtained with the application of this algorithm for two, three, and for qubit states are presented in Sec. \ref{sec:results}. Some concluding remarks are given in Sec. \ref{sec:conc}. In Appendix \ref{sec:uhlmann} we review Uhlmann's theorem and in Appendix \ref{sec:swap} we review the SWAP test. In Appendix \ref{Sec:resource}, we extend our approach to the quantification of genuine multipartite entanglement, quantum discord, quantum coherence, and total correlation. In Appendix \ref{SM:Quantum_circuit}, e provide a more detailed description of the structure of the quantum circuit used to quantify entanglement.
In Appendix \ref{SM:Additional_results}, we present additional results that complement those presented in the main text. 
In Appendix \ref{sec:comparison}, we compare our approach with other methods from the literature.

\section{Entanglement variational quantum algorithm}
\label{sec:evqa}

Variational quantum algorithms (VQAs) currently stand out as one of the leading candidates for achieving the so-called \textit{quantum advantage} --- that is, the ability of a quantum algorithm to efficiently solve problems that are intractable or difficult to address using classical methods. 
VQAs are based on an iterative procedure in which a classical optimizer adjusts the parameters $\pmb{\theta}$ of a parameterized quantum circuit $U(\pmb{\theta})$, known as the \textit{ansatz}, whose structure is defined beforehand. The goal is to find the parameters that minimize a cost function, typically defined as
\begin{equation}
C(\pmb{\theta}) = \mathrm{Tr}\left[ O U(\pmb{\theta}) \rho U(\pmb{\theta})^{\dagger} \right],
\end{equation}
where $O$ is an observable, whose choice depends on the problem at hand, and $\rho$ is the initial state of the system. The classical optimizer is usually based on the gradient descent method --- although other strategies are also explored \cite{LLES,ES_HQCNN,Anand_2021} --- and follows the update rule given by
\begin{equation}
\pmb{\theta}_{t+1} = \pmb{\theta}_{t} - \eta \nabla_{\pmb{\theta}_{t}} C(\pmb{\theta}_{t}),
\end{equation}
where $\eta$ is the learning rate, which controls how much the gradient influences the parameter update, and $t$ denotes the iteration at which the optimization is performed.

Despite their great potential, VQAs face significant challenges that still limit their large-scale practical use. One of the most critical obstacles is the phenomenon known as \textit{barren plateaus} (BPs) \cite{bp_Mcclean,bp_Wang,bp_Ortiz,bp_Cerezo,bp_Arrasmith,Friedrich_bp_2025,Ragone,Diaz}, in which the gradient of the cost function becomes exponentially small as the number of qubits increases, hindering optimization and compromising training performance. Although various approaches have been proposed to mitigate this issue \cite{bp_Friedrich,bp_Liu,bp_Liu_2}, it remains one of the main theoretical and practical challenges in the field.

Another crucial factor is the choice of \textit{ansatz}, which directly impacts the trainability and applicability of the VQA. For instance, circuits with higher expressibility, although potentially more powerful, tend to be more susceptible to the barren plateau problem and to cost function concentration \cite{Holmes_2022,Friedrich_2023}.


Given that our approach makes extensive use of Uhlmann's theorem, we give a brief overview of it, emphasizing that one of the purifications can be held fixed while only the other is varied to achieve the maximum. For more details, see Appendix \ref{sec:uhlmann}.

Let $\rho$ and $\sigma$ be two density operators acting on a Hilbert space $\mathcal{H}_A$ with dimension $d_A = \dim \mathcal{H}_A$. Then, there exist purifications $\ket{\Psi(\rho)}$ and $\ket{\Phi(\sigma)}$ of $\rho$ and $\sigma$, respectively, in an extended Hilbert space $\mathcal{H}_A \otimes \mathcal{H}_C$ of dimension $d_A d_C = \dim (\mathcal{H}_A \otimes \mathcal{H}_C)$, such that
\begin{equation}
F(\rho,\sigma) = \max_{|\Psi(\rho)\rangle,|\Phi(\sigma)\rangle}F(|\Psi(\rho)\rangle,| \Phi(\sigma)\rangle), \label{eq:uhlmann}
\end{equation}
with $F(|\Psi(\rho)\rangle,| \Phi(\sigma)\rangle) = |\langle\Psi(\rho)|\Phi(\sigma)\rangle |^2$, and the maximization runs over all possible purifications $\ket{\Psi(\rho)}$ and $\ket{\Phi(\sigma)}$. Furthermore, we may assume, without loss of generality, that $d_A = d_C$. Importantly, as noticed in Ref.~\cite{jozsa}, it is possible to maintain the purification of $\rho$ fixed while maximizing over all purification of $\sigma$, that is, 
\begin{align}
    F(\rho,\sigma) \equiv \max_{|\Phi'(\sigma)\rangle}F(|\Psi'(\rho)\rangle,|\Phi'(\sigma)\rangle), \label{eq:uhlmann1}
\end{align}
where $|\Psi'(\rho)\rangle$ is a fixed purification of $\rho$ and $|\Phi'(\sigma)\rangle$ is an arbitrary purification of $\sigma$.

\begin{figure}[t]
    \centering
    \includegraphics[width=0.8\linewidth]{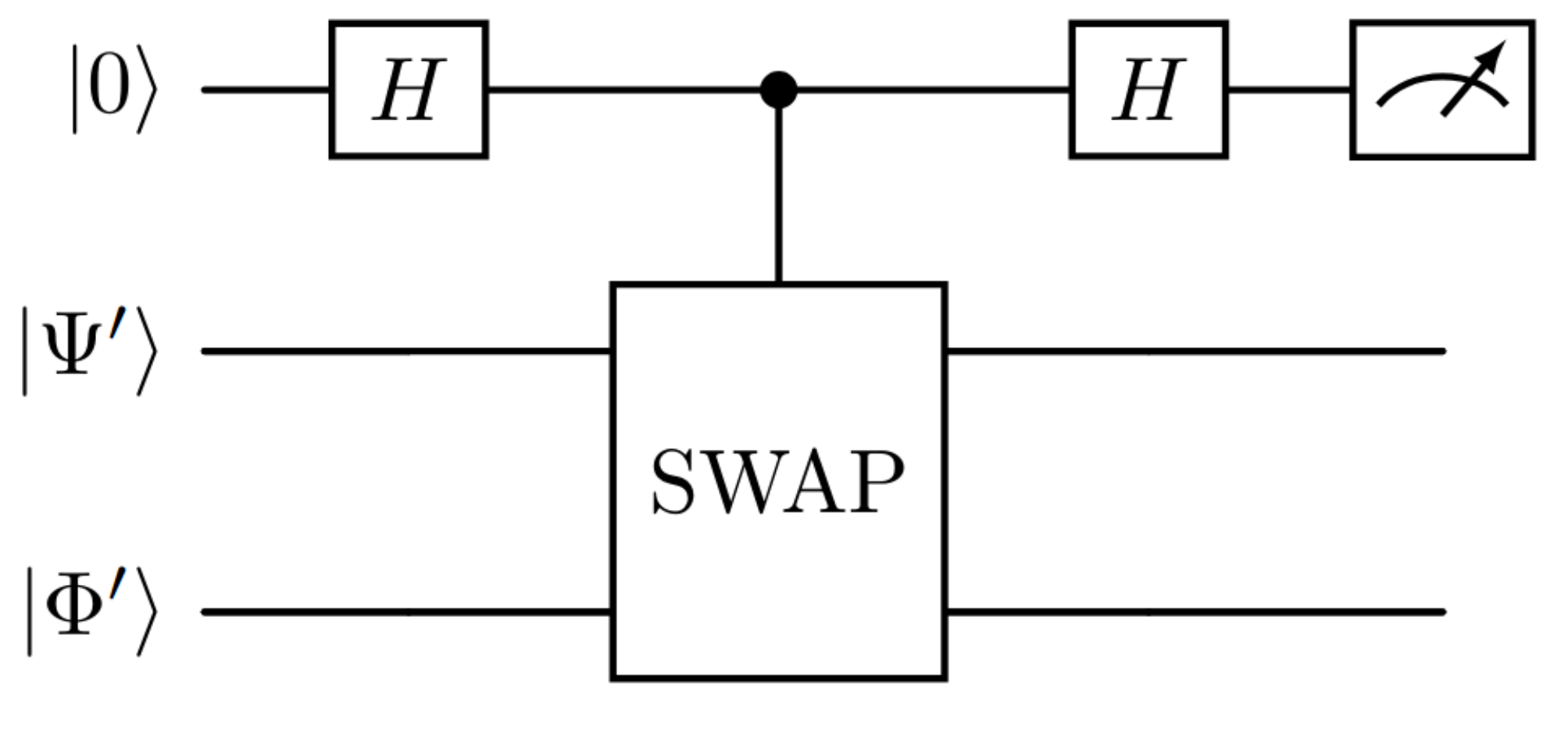}
    \caption{Illustration of the quantum circuit used to calculate the fidelity 
    between $|\Psi'\rangle := |\Psi'(\rho)\rangle  $ and $|\Phi'\rangle := |\Phi'(\sigma)\rangle $. }
    \label{fig:swapTest}
\end{figure}

The swap test, depicted in Fig. \ref{fig:swapTest}, is a quantum computing procedure used to estimate the fidelity between two states~\cite{Barenco, Buhrman, li}. Its main advantage lies in its  practicality, as it allows the evaluation of similarity between quantum states without resorting to quantum tomography, whose computational cost grows exponentially with the number of qubits in the system. The SWAP test can be straightforwardly extended to arbitrary pairs of unknown mixed quantum states $\rho$ and $\sigma$ by making use of Uhlmann's theorem.

More precisely, given that $P_0$ is the probability of obtaining the state $|0\rangle$ of the auxiliary qubit (see Appendix~\ref{sec:swap}), then the fidelity between the states $|\Psi' (\rho) \rangle$  and $|\Phi' (\sigma) \rangle$ can be expressed as
\begin{equation}
F(|\Psi'(\rho)\rangle,|\Phi'(\sigma)\rangle) = 2P_0-1.\label{eq:fidelity}
\end{equation}
Since $\ket{\Psi'(\rho)}$ is a fixed purification of $\rho$ and $\ket{\Phi'(\sigma)}$ is an arbitrary purification of $\sigma$ that has the following form
\begin{align}
    & |\Phi'(\sigma)\rangle =  (I_A\otimes U_C)|\Phi(\sigma)\rangle\label{eq:Phipuri}.
\end{align}
From Eq.~\eqref{eq:uhlmann1}, we notice that Uhlmann's theorem can be combined with the SWAP test to produce a variational quantum circuit to estimate the fidelity between any pair of mixed states $\rho$ and $\sigma$. 


We now present the protocol for estimating the Bures entanglement of arbitrary bipartite mixed quantum states, hereafter referred to as EvQA (entanglement variational quantum algorithm). Technical details are provided in Appendix~\ref{Sec:resource}. Moreover, from now on, we drop the prime and denote $|\Psi(\rho)\rangle$ as the fixed purification of $\rho$ and $|\Phi(\sigma)\rangle$ as the arbitrary purification of $\sigma$. 


The quantum entanglement $\mathcal{R}(\rho)$ contained in a state $\rho$ can be quantified from the Bures notion of distance as~\cite{Vedral97, Streltsov, Bromley}
\begin{align}
\mathcal{R}(\rho) = \min_{\sigma,|\Phi(\sigma)\rangle}2\big(1-\sqrt{F(|\Psi(\rho)\rangle,|\Phi(\sigma)\rangle)}\big) \label{eq:qresourcemain},
\end{align}
where $\sigma \equiv \rho^{AB}_{sep}$ belongs to the set of bipartite separable states, with $A$ and $B$ referring to two arbitrary partitions of a quantum system.  The minimization over $\sigma$ means that the optimization is taken over the set of bipartite separable states, while the minimization over $|\Phi(\sigma)\rangle$ is due to Uhlmann's theorem.
The state $\rho^{AB}_{sep}$ can be decomposed as a convex combination of separable pure states in the following way
\begin{align}
  \rho^{AB}_{sep} = \sum_{j=0}^{N-1}p_j|\psi_j\rangle\langle\psi_j|_A\otimes|\phi_j\rangle\langle\phi_j|_B, \label{eq:rhosepmainmain}
\end{align}
with $p_j \ge 0$ and $\sum_{j = 0}^{N-1} p_j = 1$. The parameter $N$ represents the cardinality and, in our approach, it is treated as a hyperparameter that increases from  $d_Ad_B$ up to $(d_Ad_B)^2$, with $d_A$ being the dimension of the subsystem $A$ and $d_B$ being the dimension of the subsystem $B$.

The corresponding purification of $\rho^{AB}_{sep}$ that must be optimized has the form
\begin{align}\label{eq:purificationcircuit}
|\Phi(\sigma)\rangle = U_C C^{C_j\rightarrow A}_{U_j^A}C^{C_j\rightarrow B}_{U_j^B} V_C|000\rangle_{ABC},
\end{align}
where $U_C, U_j^A, U_j^B$ and $V_C$ are arbitrary variational unitary operators. 
Moreover, $C_{U_j^A}^{C_j\rightarrow A}$ and $C_{U_j^B}^{C_j\rightarrow B}$ are controlled unitary operations of the form
\begin{align}
& C_{U_j^A}^{C_j\rightarrow A} = \sum_{j = 0}^{N-1} |j\rangle\langle j|_C\otimes U_j^A,
\end{align}
with an analogous expression for $C_{U_j^B}^{C_j\rightarrow B}$.


It is important to highlight that, during the training process, it is necessary to optimize both the bipartite separable state and its purification. Since, for each separable state, computing the fidelity requires optimizing its corresponding purification, this would, in principle, imply performing two separate training loops: one for the separable state and another for its purification. However, in practice, all parameters — both those defining the separable state and those characterizing its purification — can be optimized simultaneously. Thus, it is possible to use a single optimization loop.

\section{Results}
\label{sec:results}

To demonstrate the practical applicability proposed algorithm, we applied it to systems with two, three, and four qubits. We begin with the two-qubit case by estimating the entanglement of the well-known Werner state \cite{Werner}, which is defined as
\begin{equation}
    \rho_W = p|\Phi_+ \rangle \langle \Phi_+| + \frac{(1-p)}{4}\mathbb{I},\label{eq:wenerState}
\end{equation}
where $p \in [0,1]$, $|\Phi_+ \rangle = (|00\rangle + |11\rangle)/\sqrt{2}$, and $\mathbb{I}$ is the identity matrix $4 \times 4$. In Ref. \cite{Bromley}, the authors derived an analytical expression for the Bures measure of entanglement for the Werner state, which is known to be separable only for $p \le 1/3$.

\begin{figure}[t]
    \centering
    \includegraphics[width=0.9\linewidth]{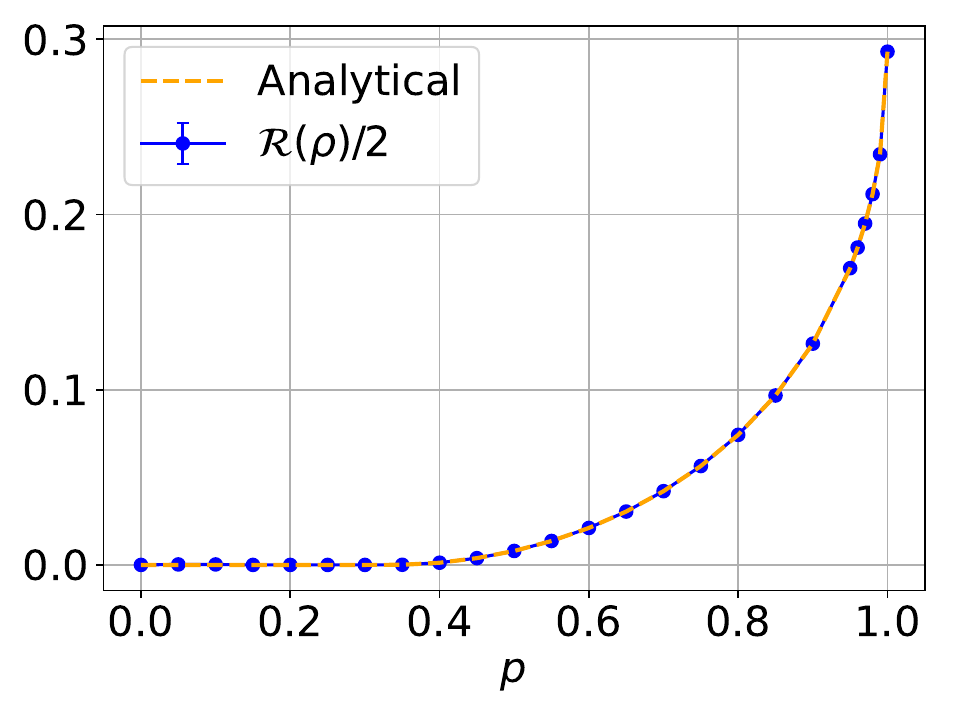}
    \caption{Comparison between the analytical expression of the Bures entanglement and EvQA results for the Werner state \ref{eq:wenerState}. The hyperparameters are given by: $N=2$ (number of qubits in subsystem $C$); $l_1=1$ (number of layers employed in the parametrization of $V_C$); $l_2=16$ (number of layers in the parametrization of $U_C$); $\eta=0.01$ (learning rate); and $1000$ Epochs.}
    \label{fig:graph_for_werner_state}
\end{figure}

In Fig. \ref{fig:graph_for_werner_state}, we present the results obtained using EvQA for the evaluation of the Bures entanglement (solid blue line with markers), along with the analytical expression of the Bures entanglement presented in Ref. \cite{Bromley} (dashed orange line). For more details on quantum circuit construction, see Appendix~\ref{SM:Quantum_circuit}. Additionally, to perform the training — both in this case and in the others — we used the Adam optimizer.

Furthermore, to analyze how the initialization of the variational parameters affects the results, we performed ten repetitions of the simulation for each value of $p$ considered, with the parameters randomly initialized in each run. Thus, in Fig. \ref{fig:graph_for_werner_state}, for each value of $p$, we plot the average value (highlighted point), along with the maximum and minimum values obtained (represented by a vertical bar). The results show that the algorithm was able to quantify the entanglement satisfactorily, since the obtained values match the analytical results. Moreover, in this specific case, the initialization of the variational parameters did not affect the results, as evidenced by the absence of visible vertical bars.

\begin{figure}[t!]
    \centering
    \includegraphics[width=1\linewidth]{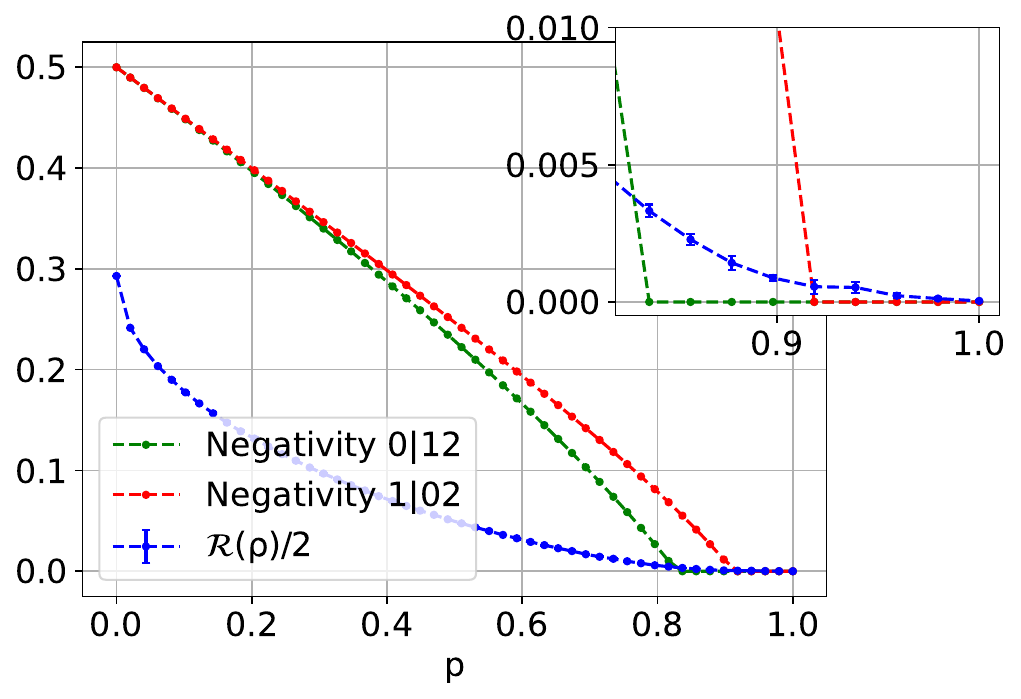}
    \caption{EvQA results for the tripartite noisy graph-state \ref{eq:lineargraphstate}. The hyperparameters are given by: $N=3$ (number of qubits in subsystem $C$); $l_1=1$ (number of layers employed in the parametrization of $V_C$); $l_2=24$ (number of layers in the parametrization of $U_C$); $\eta=0.01$ (learning rate); and $1500$ Epochs.}
    \label{fig:dephasedgraph}
\end{figure}

Moving on to our next case, we consider a tripartite qubit system, described by a linear cluster state undergoing local dephasing on each of its qubits. This state is given by
\begin{equation}
\rho_{L_3}=\left( \mathcal{E}_z \otimes \mathcal{E}_z \otimes \mathcal{E}_z \right) \left(\ket{L_3}\bra{L_3} \right),\label{eq:lineargraphstate}
\end{equation}
where $\ket{L_3}=(1/\sqrt{2})(\ket{+0+}+\ket{-1-})$ with $\ket{\pm}=(1/\sqrt{2})(\ket{0}\pm\ket{1})$ and $\mathcal{E}_z(\rho)=K_0\rho K_0  +K_1\rho K_1$ is the local dephasing channel, with $K_0=\ket{0}\bra{0}+\sqrt{1-p}\ket{1}\bra{1}$ and $K_1=\sqrt{p}\ket{1}\bra{1}$ being its Kraus operator. As shown in Fig. \ref{fig:dephasedgraph}, in the dephasing region between $0.83\leq p \leq0.91$, the state shows bound entanglement since the negativity of the bipartition $1\vert 23$ and $3\vert 12$ is zero, while the negativity of the bipartition $2|13$ is strictly positive~\cite{Cavalcanti2010,Aguilar2014}. The distance of $\rho_{L_3}$ from the closest separable state is a monotonously decreasing function of $p$, tending to zero (within the numerical precision of $10^{-3}$) as we approach $p \approx 0.91$.

\begin{figure}[t!]
    \centering
    \includegraphics[width=1\linewidth]{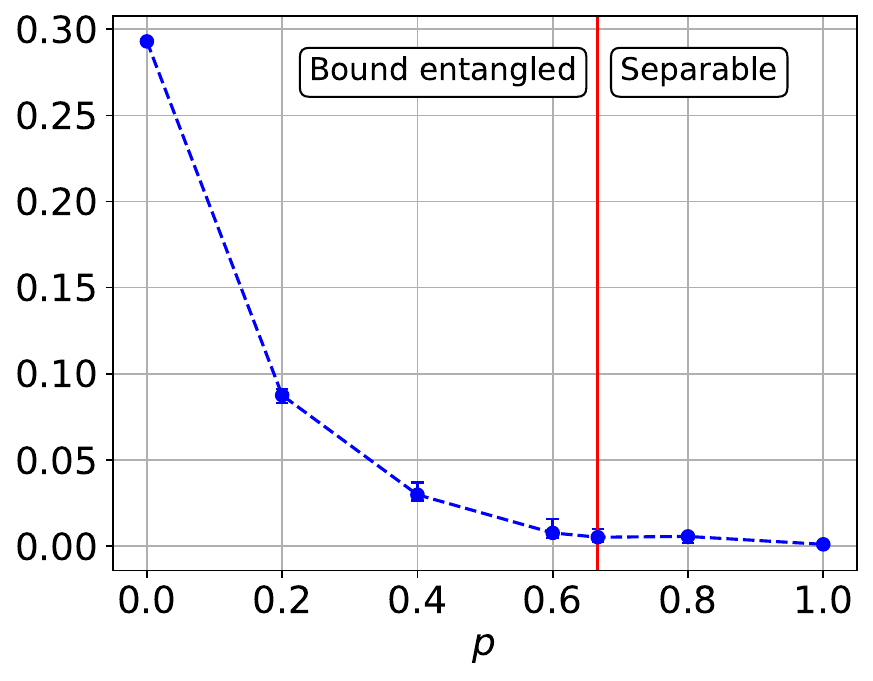}
    \caption{EvQA results for the fourpartite bound entangled Smolin state \ref{eq:noiseSmolinState}. The hyperparameters are given by: $N=5$ (number of qubits in subsystem $C$); $l_1=2$ (number of layers employed in the parametrization of $V_C$); $l_2=36$ (number of layers in the parametrization of $U_C$); $\eta=0.01$ (learning rate); and $3000$ Epochs.}
    \label{fig:graph_for_smolin_state}
\end{figure}

Finally, in Fig. \ref{fig:graph_for_smolin_state}, we consider a four-qubit system prepared in the Smolin state~\cite{Smolin}, which is defined as 
\begin{equation}
    \rho_S = (1-p)\rho_B + \frac{p}{16} \mathbb{I},\label{eq:noiseSmolinState}
\end{equation}
where $p \in [0,1]$, $\mathbb{I}$ denotes the $16 \times 16$ identity matrix,  and
\begin{equation}
    \rho_B = \frac{1}{4}\sum_{j,k=0}^1 |\Phi_{jk}\rangle\langle\Phi_{jk}|_{AB}\otimes|\Phi_{jk}\rangle\langle\Phi_{jk}|_{CD},
\end{equation}
with $|\Phi_{jk}\rangle = \big(|j\rangle\otimes|0\rangle + e^{\pi ik}|(j+1)\bmod{2}\rangle\otimes|1\rangle\big) /\sqrt{2}$ denoting all the four Bell states. This state was experimentally produced using the polarization of four optical photons, as reported in Ref.~\cite{Lavoie_2010}.  It is known to exhibit bound entanglement for $0 \le p < 2/3$, and to be fully separable for $2/3 \le p \le 1$, as shown in Fig.~\ref{fig:graph_for_smolin_state}.

\section{Concluding remarks}
\label{sec:conc}

In this article, we proposed a variational quantum algorithm to quantify quantum resources using the Bures distance, with entanglement serving as the primary figure of merit. By exploiting Uhlmann's theorem in conjunction with the SWAP test, the algorithm estimates the fidelity between a target state and a variationally optimized purification of a corresponding free state. We validated the method on two-, three-, and four-qubit systems, demonstrating accurate entanglement estimation in agreement with theoretical predictions, including cases involving bound entanglement. While minor deviations were observed near separability thresholds, these are consistent with the inherent variational and approximate nature of the approach. These observations highlight important considerations for interpreting results near critical boundaries.

First, the algorithm provides, by construction, an upper bound for the estimation of entanglement. Therefore, non-zero values obtained for $p \ge 0.91$ in Fig.~\ref{fig:dephasedgraph} and for $p \geq 2/3$ in Fig.~\ref{fig:graph_for_smolin_state} are consistent with expectations, given the approximate nature of the variational approach. However, these non-zero values remained very close to zero.

Another fundamental aspect is that the performance of the VQA is affected by the structure and expressiveness of the \textit{ansatz} $V_C$ and $U_C$ in Eq. \eqref{eq:Phipuri}. As illustrated in Fig. \ref{fig:graph_for_werner_state_using_ArbitraryUnitary} of Appendix \ref{SM:Additional_results}, the use of PennyLane's Arbitrary Unitary function, which generates highly expressive unitary matrices, for $U_C$ resulted in performances considered ideal for the analyzed problem. This suggests that the expressiveness of $U_C$ plays a crucial role in the effectiveness of the algorithm (see Appendix \ref{SM:Additional_results} for a more detailed discussion). Therefore, carefully designing or selecting an expressive $U_C$ can contribute to substantial improvements. Furthermore, the choice of hyperparameters such as learning rate, optimizer, and number of epochs also influences its performance. Hence, selecting appropriate hyperparameters is of utmost importance for the algorithm's performance. In this case, techniques such as reinforcement learning \cite{Kuo_2021} could be used to make the construction of $U_C$ and $V_C$ a learning problem and hyperparameter optimization methods \cite{Akiba_2019} could be employed to choose the most appropriate hyperparameters.

In addition, it is important to note that several well-known challenges~\cite{bp_Mcclean,bp_Wang,bp_Ortiz,bp_Cerezo,bp_Arrasmith,Friedrich_bp_2025,Holmes_2022,Friedrich_2023} still impact VQA. These issues can lead to slow convergence, requiring a large number of epochs to reach a satisfactory solution. In fact, as shown in Fig.~\ref{fig:grafico_p_variado_estado_de_smolin} of Appendix \ref{SM:Additional_results}, for certain values of $p$, convergence was notably slow, preventing the algorithm from reaching the optimal value within the number of iterations considered.

Finally, our framework extends naturally to other quantum resources, including genuine multipartite entanglement, quantum discord, coherence, and total correlations. This opens promising directions for future research, not only in the practical implementation of the algorithm for these additional resources but also in reconstructing the closest separable state in the context of entanglement.

\begin{acknowledgments}
This work was supported by the Coordination for the Improvement of Higher Education Personnel (CAPES) under Grant No. 88887.829212/2023-00, the Funda\c c\~ao de Amparo \`a Pesquisa do Estado de S\~ao Paulo (FAPESP) under Grant No.~2022/09496-8 and Grant No.~2025/07325-0, the Simons Foundation (Grant Number 1023171, RC), the Financiadora de Estudos e Projetos (grant 1699/24 IIF-FINEP) and the  National Council for Scientific and Technological Development (CNPq) under Grants No.  300083/2025-4, No. 409673/2022-6, No. 421792/2022-1, No. 307295/2020-6 and No. 403181/2024-0 and by the National Institute for the Science and Technology of Quantum Information (INCT-IQ) under Grant No.  465469/2014-0. We also thank the High-Performance Computing Center (NPAD) at UFRN for providing computational resources.
\end{acknowledgments}

\textbf{Data availability.}
The data that support the findings of this article are openly available \cite{github}.


\appendix

\section{Fidelity and Uhlmann's theorem}
\label{sec:uhlmann}

Since our method relies heavily on Uhlmann's theorem, in this section we briefly review it and highlight the fact that one of the purifications can remain fixed, while only the other has to be optimized over. Recall that the quantum fidelity between two mixed quantum states, $\rho$ and $\sigma$, is a measure of the similarity between these states defined as~\cite{nielsen}
\begin{align}
    F(\rho,\sigma) = \left(\Tr \sqrt{\sqrt{\rho}\sigma\sqrt{\rho}}\right)^2.
\end{align}
Uhlmann's theorem then states the following. Given two density operators $\rho$ and $\sigma$ defined in a Hilbert space $\mathcal{H}_A$ with dimension $d_A = \dim \mathcal{H_A}$, there exist purifications $\ket{\Psi(\rho)}$ and $\ket{\Phi(\sigma)}$ of $\rho$ and $\sigma$, respectively, in an extended Hilbert space $\mathcal{H}_A \otimes \mathcal{H}_C$ with dimension $d_A d_C = \dim (\mathcal{H}_A \otimes \mathcal{H}_C)$ such that
\begin{align}
F(\rho,\sigma) & = \max_{|\Psi(\rho)\rangle,|\Phi(\sigma)\rangle}F(|\Psi(\rho)\rangle,|\Phi(\sigma)\rangle) \\ & = \max_{|\Psi(\rho)\rangle,|\Phi(\sigma)\rangle} |\langle\Psi(\rho)|\Phi(\sigma)\rangle |^2 ,
\end{align}
where the maximization is taken over all possible purifications $\ket{\Psi(\rho)}$ and $\ket{\Phi(\sigma)}$. Moreover, without loss of generality, we assume that $d_A = d_C$.

Hence, given the spectral decomposition of the density operators
\begin{align}
\rho = \sum_{j = 0}^{d_A -1} p_j|\psi_j\rangle_A\langle \psi_j|, \ \ \ \sigma = \sum_{j = 0}^{d_A -1} q_j|\phi_j\rangle_A\langle \phi_j|,
\end{align}
their respective purifications can be written as
\begin{align}
& |\Psi(\rho)\rangle = \sum_{j = 0}^{d_A -1} \sqrt{p_j}|\psi_j\rangle_A\otimes|b_j\rangle_C, \\ & |\Phi (\sigma)\rangle = \sum_{j = 0}^{d_A -1} \sqrt{q_j}|\phi_j\rangle_A\otimes|c_j\rangle_C,
\end{align}
where the orthonormal basis $|b_j\rangle_C$ and $|c_j\rangle_C$ can be obtained from the computational
basis $\ket{j}_C$. That is, $|b_j\rangle_C = U|j\rangle_C$ and $|c_j\rangle_C = V|j\rangle_C$ with $U$ and $V$ being unitary operators. Now, by noticing that
\begin{align}
& \langle\Psi(\rho)|\Phi(\sigma)\rangle \nonumber \\
& = \sum_{j = 0}^{d_A -1}\sqrt{p_j}\langle \psi_j|_A\otimes\langle b_j|_C\sum_{k = 0}^{d_A -1} \sqrt{q_k}|\phi_k\rangle_A\otimes|c_k\rangle_C \nonumber \\ 
& = \sum_{j = 0}^{d_A -1}\sqrt{p_j}\langle \psi_j|_A\otimes\langle j|_C\sum_{k = 0}^{d_A -1} \sqrt{q_k}|\phi_k\rangle_A\otimes|d_k\rangle_C \\ 
& =: \langle\Psi'(\rho)|\Phi'(\sigma)\rangle\nonumber,
\end{align}
where $|d_k\rangle_C \equiv U^{\dagger} V \ket{k}_C$ is also an orthonormal basis and we defined

\begin{equation}
    |\Psi'(\rho)\rangle \equiv \sum_{j = 0}^{d_A -1}\sqrt{p_j}|\psi_j\rangle_A\otimes|j\rangle_C 
\end{equation}
and
\begin{equation}
    \quad |\Phi'(\sigma)\rangle \equiv \sum_{k = 0}^{d_A -1} \sqrt{q_k}|\phi_k\rangle_A\otimes|d_k\rangle_C. \label{eq:purifications_of_rho_and_sigma}
\end{equation}


This implies that we can maintain the purification of $\rho$ fixed while maximizing over all purifications of $\sigma$, such that
\begin{align}
    F(\rho,\sigma) & = \max_{|\Psi(\rho)\rangle,|\Phi(\sigma)\rangle}F(|\Psi(\rho)\rangle,|\Phi(\sigma)\rangle) \\ & = \max_{|\Phi'(\sigma)\rangle}F(|\Psi'(\rho)\rangle,|\Phi'(\sigma)\rangle) \label{Ffix}, 
\end{align}
as noticed in Ref.~\cite{jozsa}.

\section{Variational SWAP test for mixed state fidelity estimation}
\label{sec:swap}

The SWAP test can be used to calculate the fidelity between any pair of unknown pure quantum states of qudits by estimating the probabilities of measurements on an auxiliary qubit.

Specifically, given two pure states \( |\Phi\rangle \) and \( |\Psi\rangle \), each defined in a Hilbert space of dimension \( d = 2^n \), where \( n \) is the number of qubits, the fidelity between them can be estimated through a quantum circuit with \( 2n + 1 \) qubits. Initially, the system is prepared in the state \( |0\rangle \otimes |\Phi\rangle \otimes |\Psi\rangle \). A Hadamard gate is then applied to the first qubit, resulting in the state
\begin{equation}
    \frac{1}{\sqrt{2}} \left( |0\rangle \otimes |\Phi\rangle \otimes |\Psi\rangle + |1\rangle \otimes |\Phi\rangle \otimes |\Psi\rangle \right).
\end{equation}
A controlled-swap gate is then applied, with the first qubit as control, yielding the state
\begin{equation}
    \frac{1}{\sqrt{2}} \left( |0\rangle \otimes |\Phi\rangle \otimes |\Psi\rangle + |1\rangle \otimes |\Psi\rangle \otimes |\Phi\rangle \right).
\end{equation}
After applying a second Hadamard gate to the first qubit, the system evolves into
\begin{equation}
    \frac{1}{2} |0\rangle \otimes (|\Phi\rangle \otimes |\Psi\rangle + |\Psi\rangle \otimes |\Phi\rangle) + \frac{1}{2} |1\rangle \otimes (|\Phi\rangle \otimes |\Psi\rangle - |\Psi\rangle \otimes |\Phi\rangle).
\end{equation}
Measuring the first qubit, the probability $P_0$ of obtaining the state \( |0\rangle \) of the auxiliary qubit is given by
\begin{equation}
    P_0 = \frac{1}{2} + \frac{1}{2} |\langle \Psi | \Phi \rangle|^2.
\end{equation}
Therefore, the fidelity between the states can be obtained using the expression
\begin{equation}
    F(|\Phi\rangle, |\Psi\rangle) = 2P_0 - 1.
\end{equation}

Now, the SWAP test can be easily generalized for any pair of unknown mixed quantum states $\rho$ and $\sigma$ by using Uhlmann's theorem described in Appendix \ref{sec:uhlmann}. Given a fixed purification $\ket{\Psi'(\rho)}$ of $\rho$ and an arbitrary purification $\ket{\Phi'(\sigma)}$ of $\sigma$,
we observe from Eq.~\eqref{Ffix} that Uhlmann's theorem can be leveraged alongside the SWAP test to construct a variational quantum circuit capable of estimating the fidelity between any two mixed quantum states.

\section{Bures resources quantification}
\label{Sec:resource}

Following Ref.~\cite{Vedral97, Streltsov, Bromley}, by using the Bures notion of distance defined on the set of mixed quantum states, we notice that quantum resources can be quantified in a universal way. Let us then begin by defining the Bures distance as
\begin{align}
    \mathcal{D}^2_B(\rho, \sigma) =  2\big(1-\sqrt{F(\rho,\sigma)}\big), \label{eq:squaredbures}
\end{align}
where $\rho$ and $\sigma$ are two arbitrary states. It is worth mentioning that $\mathcal{D}^2_B$ is, in fact, the squared Bures distance~\cite{Adesso}. However, in this work we do not make this distinction. For instance, Refs.~\cite{Vedral97, Streltsov} refer to Eq.~\eqref{eq:squaredbures} as the Bures distance, while Ref.~\cite{Adesso} refers to it as the squared Bures distance.

Consider now a given quantum resource $\mathcal{R}(\rho)$ of state $\rho$, such as entanglement, quantum discord, or quantum coherence. Let $\sigma$ be a free state of this quantum resource, such as a separable state, a classical-quantum state, or an incoherent state. Then the quantum resource contained in the state $\rho$ can be quantified as
\begin{align}
\mathcal{R}(\rho) & = \min_{\sigma}\mathcal{D}^2_B(\rho,\sigma) \nonumber \\
& = \min_{\sigma}2\big(1-\sqrt{\max_{|\Phi(\sigma)\rangle}F(|\Psi(\rho)\rangle,|\Psi(\sigma)\rangle)}\big) \nonumber \\
& = \min_{\sigma,|\Phi(\sigma)\rangle}2\big(1-\sqrt{F(|\Psi(\rho)\rangle,|\Phi(\sigma)\rangle)}\big) \label{eq:qresource},
\end{align}
where the minimization over $\sigma$ means that the optimization is taken over the set of all free states, while the minimization over $|\Phi(\sigma)\rangle$ is due to Uhlmann's theorem as discussed in the Appendix \ref{sec:uhlmann}.

\subsection{Entanglement quantification}
If the set of free states is the set of mixed bipartite separable states, then the quantum resource being quantified in Eq.~\eqref{eq:qresource} is entanglement.

Given that a bipartite separable state $\sigma \equiv \rho^{AB}_{sep}$, where $A$ and $B$ refer to two arbitrary partitions of a quantum system, can be decomposed as a convex combination of separable pure states in the following way
\begin{align}
  \rho^{AB}_{sep} = \sum_{j=0}^{N-1}p_j|\psi_j\rangle\langle\psi_j|_A\otimes|\phi_j\rangle\langle\phi_j|_B, \label{eq:rhosep}
\end{align}
with $p_j \ge 0$ and $\sum_{j = 0}^{N=1} p_j = 1$. 

The corresponding purification of $\rho^{AB}_{sep}$ that must be optimized has the form
\begin{align}
|\Phi(\rho^{AB}_{sep})\rangle & = \sum_{j=0}^{N-1}\sqrt{p_j}|\psi_j\rangle_A\otimes|\phi_j\rangle_B\otimes|c_j\rangle_C \nonumber \\
& = \sum_{j=0}^{N-1}\sqrt{p_j}U_j^A|0\rangle_A\otimes U_j^B|0\rangle_B\otimes U_C|j\rangle_C \nonumber \\
& = U_C C^{C_j\rightarrow A}_{U_j^A}C^{C_j\rightarrow B}_{U_j^B}|0\rangle_A\otimes |0\rangle_B\otimes \sum_{j=0}^{N-1}\sqrt{p_j} |j\rangle_C \nonumber \\
& = U_C C^{C_j\rightarrow A}_{U_j^A}C^{C_j\rightarrow B}_{U_j^B} V_C|000\rangle_{ABC}, \label{eq:rho_sep_AB_purification}
\end{align}
with $U_C, U_j^A, U_j^B$ and $V_C$ being arbitrary variational unitary operators.

Initially, the unitary $V_C$ must be constrained to generate a superposition of states with real coefficients. This constraint arises from the fact that $V_C$ is associated with the generation of the state
\[
\sum_{j=0}^{N-1} \sqrt{p_j} \, |j\rangle_C,
\]
which, in turn, is linked to the probabilities \( \{p_j\}_{j=0}^{N-1} \) in Eq.~\eqref{eq:rhosep}. However, upon noticing that these probabilities can be expressed as
\[
p_j = |c_j|^2, \quad \forall j, \quad \text{with } c_j \in \mathbb{C},
\]
it follows that \( V_C \) no longer needs to be restricted, becoming a more general unitary.

In addition, the operators 

\begin{equation}
    C_{U_j^A}^{C_j\rightarrow A} = \sum_{j = 0}^{N-1} |j\rangle\langle j|_C\otimes U_j^A
\end{equation}
and
\begin{equation}
    C_{U_j^B}^{C_j\rightarrow B} = \sum_{j = 0}^{N-1} |j\rangle\langle j|_C\otimes U_j^B,
\end{equation}
are controlled unitaries, where the control is on register $C$ and the targets are $A$ and $B$.

It is worth mentioning that, in Eq.~\eqref{eq:rhosep}, the parameter $N$ is the cardinality (the number of pure states needed in the convex mixture), with $\text{rank}(\rho^{AB}_{sep}) \le N \le (d_A d_B)^2$~\cite{Sanpera, Horodecki1, Lockart}. In particular, for 2-qubit states, it is possible to show that $N\le 4$. In our method, we treat cardinality as a hyperparameter and increase it from $d_A d_B$ to $(d_A d_B)^2$ to see if a tighter upper bound exists.

\begin{figure}[ht]
    \centering
    \includegraphics[width=1\linewidth]{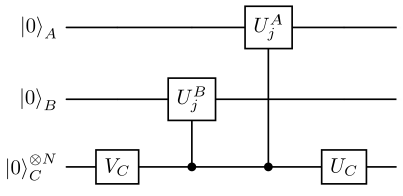}
    \caption{Illustration of the circuit used for the preparation of the $|\Psi(\rho_{sep}^{AB})\rangle$ state from Eq. \eqref{eq:rho_sep_AB_purification}.}
    \label{fig:Phi_purification}
\end{figure}

Finally, another point worth noting is that our method allows for the reconstruction of the closest separable states. Once the VQA training is done, we obtain $U_C$. Hence, we can apply
\begin{align}
U_C^\dagger|\Psi(\rho^{AB}_{sep})\rangle = \sum_{j=0}^{N-1}\sqrt{p_j}|\psi_j\rangle_A\otimes|\phi_j\rangle_B\otimes|j\rangle_C.
\end{align}
Now, measuring $C$ on the computational basis, we reconstruct the probabilities $p_j = \Pr(|j\rangle_C).$ By post-selecting $|j\rangle_C$ together with the measurement of a complete set of observables for the subsystems $A$ and $B$, we can reconstruct the states $|\psi_j\rangle_A$ and $|\phi_j\rangle_B$. With this, we obtain information about the whole structure of $\rho^{AB}_{sep}$.
The variational quantum circuit to prepare the purification of $\rho^{AB}_{sep}$ is illustrated in Fig. \ref{fig:Phi_purification}.

\subsection{Genuine multipartite entanglement quantification}\label{subsec:gmeq}
Interestingly, our method also applies to the quantification of genuine multipartite entanglement. If the set of free states is the set of bi-separable states, then the quantum resource being quantified in Eq.~\eqref{eq:qresource} is genuine multipartite entanglement. 

Let us then define a biseparable state following the Refs.~\cite{Huber, Palazuelos}. Given a fixed bi-partition $\mathcal{A} = \{\mathcal{B}, \bar{\mathcal{B}}\}$ of a multipartite quantum system of $M$ parts, with $\bar{\mathcal{B}}$ being the complement of $\mathcal{B}$, a separable state with fixed bi-partition is given by
\begin{align}
  \rho_{\mathcal{B}\bar{\mathcal{B}}} = \sum_{j = 0}^{N-1} p_j |\psi_j\rangle_{\mathcal{B}}\langle \psi_j| \otimes |\phi_j\rangle_{\bar{\mathcal{B}}}\langle \phi_j|. \label{eq:sepa} 
\end{align}
A bi-separable state is obtained by considering convex combinations of the different bi-partitions from the Eq.~\eqref{eq:sepa}, i.e.,
\begin{align}
    \rho_{\text{bi-sep}} = \sum_{k_\mathcal{B}} q_{k_\mathcal{B}} \left(\sum_{j = 0}^{N-1} p^{k_\mathcal{B}}_j |\psi_j\rangle_{\mathcal{B}}\langle \psi_j| \otimes |\phi_j\rangle_{\bar{\mathcal{B}}}\langle \phi_j| \right),\label{eq:bi-sep}
\end{align}
where the index $k_{\mathcal{B}}$ enumerates the different possible bi-partitions, $q_{k_\mathcal{B}}  \ge 0$ and  $\sum_{k_\mathcal{B}} q_{k_\mathcal{B}} = 1$. It is worth mentioning that the sum over $k_{\mathcal{B}}$ will contain at most $2^{M-1} - 1$ terms. Since the upper bound of this sum is not fixed, we chose not to make it explicit in the Eq.~\eqref{eq:bi-sep}.

\begin{figure}[h!]
    \centering
    \includegraphics[width=0.8\linewidth]{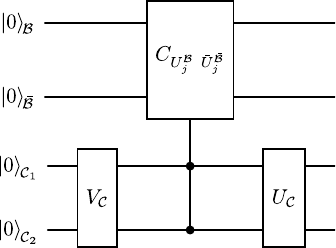}
    \caption{Here we show the variational quantum circuit to prepare the state described by Eq. \eqref{eq:ansatz_gm}.}
    \label{fig:placeholder}
\end{figure}

The corresponding purification of $\rho_{\text{bi-sep}}$ that must be optimized has the form
\begin{align}\label{eq:ansatz_gm}
& |\Phi(\rho_{\text{bi-sep}})\rangle  = \sum_{k_\mathcal{B}}\sum_{j =0}^{N-1} \sqrt{q_{\mathcal{B}} p^{k_\mathcal{B}}_j} |\psi_j\rangle_{\mathcal{B}}  |\phi_j\rangle_{\bar{\mathcal{B}}}  |c_{jk_\mathcal{B}}\rangle_{\mathcal{C}}\nonumber \\
& =  U_{\mathcal{C}} \sum_{k_\mathcal{B}}\sum_{j =0}^{N-1} \sqrt{q_{\mathcal{B}} p^{k_\mathcal{B}}_j} U_j^{\mathcal{B}} |j\rangle_{\mathcal{B}}  \bar{U}_j^{\bar{\mathcal{B}}} |j\rangle_{\bar{\mathcal{B}}}  |j\rangle_{\mathcal{C}_1}  |k_\mathcal{B}\rangle_{\mathcal{C}_2}\\
& =  U_{\mathcal{C}} C_{U_j^{\mathcal{B}} \ \bar{U}_j^{\bar{\mathcal{B}}}}^{\mathcal{C} \to \mathcal{B},\bar{\mathcal{B}}} V_{\mathcal{C}} |0\rangle_{\mathcal{B}}  |0\rangle_{\bar{\mathcal{B}}}  |0\rangle_{\mathcal{C}_1} |0\rangle_{\mathcal{C}_2}, \nonumber 
\end{align} 
where
\begin{equation}
C_{U_j^{\mathcal{B}} \ \bar{U}_j^{\bar{\mathcal{B}}}}^{\mathcal{C} \to \mathcal{B},\bar{\mathcal{B}}} = \sum_{k_{\mathcal{B}}}\sum_{j =0}^{N-1} |j\rangle_{\mathcal{C}_1} \langle j | \otimes |k_\mathcal{B}\rangle_{\mathcal{C}_2} \langle k_\mathcal{B}| \otimes U_j^{\mathcal{B}} \otimes \bar{U}_j^{\bar{\mathcal{B}}} ,
\end{equation}
with $ U_{\mathcal{C}} , U_j^{\mathcal{B}},  \bar{U}_j^{\bar{\mathcal{B}}}, V_{\mathcal{C}} $ being arbitrary variational unitary operators. It is worth mentioning that $U_j^{\mathcal{B}}$ and $\bar{U}_j^{\bar{\mathcal{B}}}$ acts on the partitions $\mathcal{B}$ and $\bar{\mathcal{B}}$, respectively, which are identified by the index $k_{\mathcal{B}}$ and the auxiliary system $\mathcal{C}$ is composed of two qudits $\mathcal{C}_1$ and $\mathcal{C}_2$. Moreover, $V_{\mathcal{C}}$ prepares an arbitrary superposition state with real coefficients.
The variational quantum circuit to prepare the purification of $\rho_{\text{bi-sep}}$ is illustrated in Fig. \ref{fig:placeholder}.

\subsection{Quantum discord quantification}\label{subsec:qdq}
If the set of free states is the set of quantum-classical states, then the quantum resource being quantified in Eq.~\eqref{eq:qresource} is quantum discord.
A bipartite quantum-classical state has the form
\begin{align}
\rho_{qc}^{AB} & = \sum_{j=0}^{d_B-1}p_{j}\rho_j^A\otimes|\phi_j\rangle\langle \phi_j|_B \nonumber \\
& = \sum_{j=0}^{d_B-1}\sum_{k=0}^{d_A-1}p_{j}\alpha_{jk}|\psi_{jk}\rangle\langle \psi_{jk}|_A\otimes|\phi_j\rangle\langle \phi_j|_B,
\end{align}
where we used the spectral decomposition for $d_B$ density operators $\rho_j^A $, $\alpha_{jk} \ge 0$, and $\sum_{k =0}^{d_A -1} \alpha_{jk} = 1$.

\begin{figure}[h]
    \centering
    \includegraphics[width=1\linewidth]{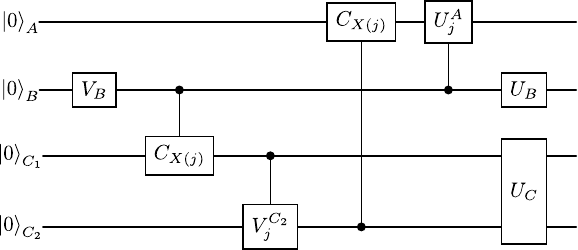}
    \caption{This figure shows the variational quantum circuit to prepare the purification of the state $\rho_{qc}^{AB}$ described by Eq.~\eqref{eq:ansatz_discord}.}
\label{fig:ansatz_discord}
\end{figure}

The corresponding purification of $\rho_{qc}^{AB}$ that must be optimized has the form
\begin{align}
& |\Phi(\rho^{AB}_{qc})\rangle = \sum_{j=0}^{d_B-1}\sum_{k=0}^{d_A-1}\sqrt{p_{j}\alpha_{jk}}|\psi_{jk}\rangle_A|\phi_j\rangle_B|c_{jk}\rangle_C \nonumber \\
& = \sum_{j=0}^{d_B-1}\sum_{k=0}^{d_A-1}\sqrt{p_{j}\alpha_{jk}}U_j^A|k\rangle_A U_B|j\rangle_B U_C|jk\rangle_C \nonumber \\
& = U\sum_{j=0}^{d_B-1}\sqrt{p_{j}}U_j^A C_{X(j)}^{(C_2)_j\rightarrow A}|0\rangle_A|j\rangle_B |j\rangle_{C_1} \sum_{k=0}^{d_A-1}\sqrt{\alpha_{jk}}|k\rangle_{C_2} \nonumber \\
& = U\sum_{j=0}^{d_B-1}\sqrt{p_{j}}U_j^A C_{X(j)}^{(C_2)_j\rightarrow A}|0\rangle_A|j\rangle_B |j\rangle_C V_j^{C_2}|0\rangle_{C_2} \nonumber \\
& = U W|0\rangle_A\otimes\sum_{j=0}^{d_B-1}\sqrt{p_{j}}|j\rangle_B |j\rangle_{C_1}|0\rangle_{C_2} \nonumber \\
& = U WC_{X(j)}^{B_j\rightarrow C_1}|0\rangle_A\otimes\sum_{j=0}^{d_B-1}\sqrt{p_{j}}|j\rangle_B\otimes |0\rangle_{C_1}\otimes|0\rangle_{C_2}\nonumber \\
& = U WC_{X(j)}^{B_j\rightarrow C_1}V_B|0000\rangle_{ABC_1 C_2},\label{eq:ansatz_discord}
\end{align}
where $U = (\mathbb{I}_A \otimes U_B \otimes U_C)$, $W = C_{U_j^A}^{B_j\rightarrow A} C_{X(j)}^{(C_2)_j\rightarrow A}C_{V_j^{C_2}}^{(C_1)_j\rightarrow C_2}$ and
\begin{align}
& C_{X(j)}^{(C_2)_j\rightarrow A} = \sum_{j = 0}^{d_A - 1} |j\rangle\langle j|_{C_2}\otimes X(j)_A,
\end{align}
with the state shift operator defined by $X(j)|k\rangle = |(j+k)\bmod d_A\rangle$ and similarly for $C_{X(j)}^{B_j\rightarrow C_1}$. The variational quantum circuit to prepare the purification of $\rho^{AB}_{qc}$ is illustrated in Fig. \ref{fig:ansatz_discord}.

\begin{figure}[h]
    \centering
    \includegraphics[width=0.6\linewidth]{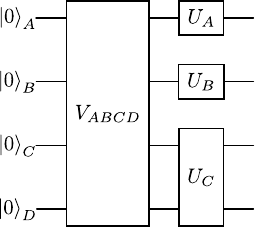}
    \caption{Illustration of the variational quantum circuit used to prepare the purification of $\rho^{AB}_{cc}$ defined in Eq.~\eqref{eq:ansatz_classical_classical}.}
    \label{fig:ansatz_classical_classical}
\end{figure}

If quantum discord is quantified with relation to classical-classical states
$$\rho_{cc}^{AB} = \sum_{j=0}^{d_A-1}\sum_{k=0}^{d_B-1}p_{jk}|a_j\rangle\langle a_j|_A\otimes|b_k\rangle\langle b_k|_B,$$
with $p_{jk}$ being a probability distribution and $\{|a_j\rangle\},\{|b_k\rangle\}$ are orthonormal bases, the purification to be optimized is
\begin{align}
|\Psi(\rho^{AB}_{cc})\rangle 
& = \sum_{j=0}^{d_A-1}\sum_{k=0}^{d_B-1}\sqrt{p_{jk}}|a_j\rangle_A\otimes|b_k\rangle_B\otimes|c_{jk}\rangle_C \\
& = \sum_{j=0}^{d_A-1}\sum_{k=0}^{d_B-1}\sqrt{p_{jk}}U_A|j\rangle_A\otimes U_B|k\rangle_B\otimes U_C|jk\rangle_C \\
& = U_{ABC}\sum_{j=0}^{d_A-1}\sum_{k=0}^{d_B-1}\sqrt{p_{jk}}|j\rangle_A\otimes|k\rangle_B\otimes |jk\rangle_C \\
& = U_{ABC}|\Psi_f\rangle_{ABCD}, \label{eq:ansatz_classical_classical}
\end{align}
where $\{|a_j\rangle\},\ \{|b_j\rangle\},\ \{|c_j\rangle\}$ are orthonormal bases, $U_{ABC} = (U_A\otimes U_B\otimes U_C)$, $U_S,\ S=A,B$, are general one-qudit unitary transformations, $U_C$ is a general two-qudit unitary transformation, and we defined the purification
\begin{align}
|\Psi_f\rangle_{ABCD} & = \sum_{j=0}^{d_A-1}\sum_{k=0}^{d_B-1}\sqrt{p_{jk}}|j\rangle_A\otimes|k\rangle_B\otimes |jk\rangle_C \\
& = V_{ABCD}|0000\rangle_{ABCD},
\end{align}
that depends only on the probability distribution $\{p_{jk}\}.$ The variational quantum circuit to prepare the purification of $\rho^{AB}_{cc}$ is illustrated in Fig. \ref{fig:ansatz_classical_classical}.

\subsection{Quantum coherence quantification}\label{subsec:qcq}
If the set of free states is the set of incoherent states, then the quantum resource being quantified in Eq.~\eqref{eq:qresource} is quantum coherence.

An incoherent state has the form 
\begin{align}
    \rho_{\iota}^A= \sum_{j=0}^{d_A-1}p_j|j\rangle_A\langle j|
\end{align}
with the basis $\{|j\rangle\}$ fixed.

\begin{figure}[h]
    \centering
    \includegraphics[width=0.6\linewidth]{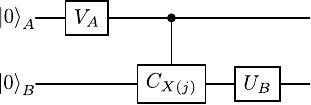}
    \caption{In this figure we show the variational quantum circuit to prepare purifications of the state $\rho_{\iota}^A$, described by Eq.~\eqref{eq:ansatez_coherence}. }
    \label{fig:ansatez_coherence}
\end{figure}

The corresponding purification of $\rho_{\iota}^A$ that must be optimized has the form
\begin{align}
|\Phi(\rho_{\iota}^A)\rangle & = \sum_{j=0}^{d-1}\sqrt{p_j}|j\rangle_A\otimes|\phi_j\rangle_B \nonumber \\
& = U_B C_{X(j)}^{A_j\rightarrow B} V_A|0\rangle_A\otimes|0\rangle_B.\label{eq:ansatez_coherence}
\end{align}
where $U_B$ and $V_A$ are general unitary. 
The variational quantum circuit to prepare the purification of $\rho_\iota^A$ is illustrated in Fig. \ref{fig:ansatez_coherence}.

\subsection{Total correlation quantification}\label{subsec:tcq}
If the set of free states is the set of mixed product states, then the resource being quantified in Eq.~\eqref{eq:qresource} is the total correlations, which can be a mixture of quantum and classical correlations.

A mixed product state has the form
\begin{align}
    \rho^{AB}_{p} = \rho^A  \otimes \rho^B = \sum_{j = 0}^{d_A-1} \sum_{k = 0}^{d_B-1}p_j q_k|\psi_j\rangle\langle \psi_j|\otimes|\phi_k\rangle\langle \phi_k|.
\end{align}

\begin{figure}[ht]
    \centering
    \includegraphics[width=0.8\linewidth]{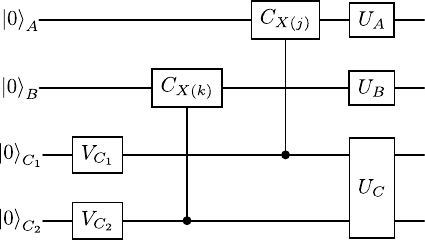}
    \caption{Here we show the variational quantum circuit to prepare the purification described by Eq.~\eqref{eq:anstaz_total_correlation}.}
    \label{fig:ansatz_total_correlation}
\end{figure}

The corresponding purification of $\rho^{AB}_{p}$ that must be optimized has the form
\begin{align}
|\Phi(\rho^{AB}_{p})\rangle_{AB} & =\sum_{j = 0}^{d_A-1} \sum_{k = 0}^{d_B-1}\sqrt{p_j q_k}|\psi_j\rangle_A\otimes|\phi_k\rangle_B\otimes|c_{jk}\rangle_C \nonumber \\ 
& = U_A U_B U_C C^{AB}V_{C_1}V_{C_2}|0000\rangle_{ABC_1 C_2},\label{eq:anstaz_total_correlation}
\end{align}
with
\begin{equation}
   C^{AB} =  C_{X(j)}^{(C_1)_j\rightarrow A}C_{X(k)}^{(C_2)_k\rightarrow B}.
\end{equation}
Here, $U_A,U_B,U_C$ are general variational forms and $V_{C_1}, V_{C_2}$ are arbitrary variational forms that prepare states with $p_j = |c_j|^2$ and $q_k = |d_k|^2$, where $c_j, d_k \in \mathbb{C}$. 
The variational quantum circuit to prepare the purification of $\rho_p^{AB}$ is illustrated in Fig. \ref{fig:ansatz_total_correlation}.

\section{Quantum circuit}\label{SM:Quantum_circuit}

In this appendix, we provide a more detailed description of the circuit structure used to quantify entanglement, specifically the fidelity between the states $|\Psi(\rho)\rangle$ and $|\Phi(\sigma)\rangle$, as defined in Eq. \eqref{eq:qresourcemain}. The purification $|\Psi(\rho)\rangle$ depends on the state $\rho$ under consideration, and in some cases, this purification is already known — for example, for two-qubit Bell-diagonal states \cite{Pozzobom_2019}. Since the objective of this work is to employ the same circuit regardless of the state, we adopt the purification described in Ref. \cite{Pinto_2024}.

Consider an arbitrary $n$-qubit state given by
\begin{equation}
  \rho = \sum_{j=0}^{2^n-1} r_j |\phi_j \rangle \langle \phi_j|.  
\end{equation}
The purification of this state, according to Ref. \cite{Pinto_2024}, is obtained by the following procedure. Initially, a circuit with $2n$ qubits is prepared. Then, a parameterization $V_{|0\rangle \rightarrow |\psi\rangle}$ is applied to the first $n$ qubits, whose purpose is to prepare the state
\begin{equation}
    |\psi\rangle = \sum_{j=0}^{2^n-1} \sqrt{r_j}|j\rangle |0\rangle^{\otimes n}.
\end{equation}
Next, a sequence of CNOT gates is applied between the first $n$ qubits and the remaining $n$ qubits, with the former acting as control qubits and the latter as targets. This generates the state
\begin{equation}
    |\psi'\rangle = \sum_{j=0}^{2^n-1} \sqrt{r_j}|j\rangle |j\rangle.
\end{equation}
Finally, a second parametrization $U_{ |j\rangle \rightarrow |\phi_j\rangle }$ is applied, yielding the purification of $\rho$, that is
\begin{equation}
    |\Psi(\rho)\rangle = \sum_{j=0}^{2^n-1} \sqrt{r_j}|\phi_j\rangle |j\rangle.\label{eq:rho_purif}
\end{equation}

\begin{figure}[ht]
    \centering
    \includegraphics[width=0.8\linewidth]{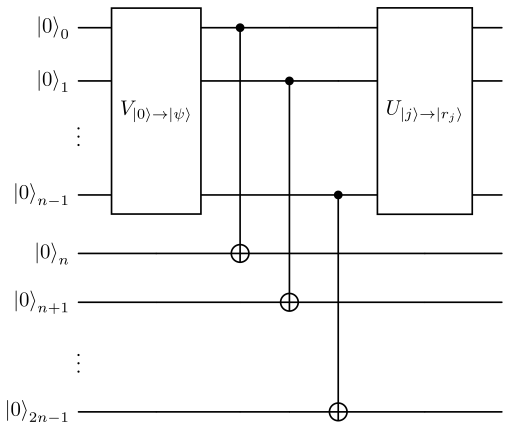}
    \caption{Illustration of the circuit used for the preparation of the $|\Psi(\rho)\rangle$ state from Eq. \eqref{eq:rho_purif}.}
    \label{fig:dataEncoder}
\end{figure}

Fig. \ref{fig:dataEncoder} shows an illustration of the circuit used to generate this purification. Since it is valid for any state $\rho$, this circuit is employed to generate the purification for the three states described in Eqs. \eqref{eq:wenerState}, \eqref{eq:lineargraphstate}, and \eqref{eq:noiseSmolinState}, which are analyzed in this work.

Furthermore, it should be noted that the procedure used to obtain the purification $|\Psi(\rho)\rangle$, as described in Ref. \cite{Pinto_2024}, applies not only to qubits but also to qudits. Since the proposed entanglement quantification algorithm is likewise applicable to qudits, the circuit presented in Fig. \ref{fig:dataEncoder} can be adopted as a standard routine for its implementation.

The purification corresponding to the state $|\Phi(\sigma)\rangle = |\Psi(\rho_{sep}^{AB})\rangle$ was previously obtained in Eq.~\eqref{eq:rho_sep_AB_purification}. Fig.~\ref{fig:Phi_purification} presents the quantum circuit used to prepare this purification. Although the figure illustrates the preparation of the state $|\Phi(\sigma)\rangle = |\Psi(\rho_{sep}^{AB})\rangle$ for a bipartite system, the generalization to the cases presented in Eq.~\eqref{eq:lineargraphstate} and Eq.~\eqref{eq:noiseSmolinState} is straightforward.

As shown in Fig.~\ref{fig:Phi_purification}, the circuit begins with the three subsystems $A$, $B$, and $C$ initialized in the state $|0\rangle$. A general unitary $V_C$ is then applied to subsystem $C$, followed by the application of multi-controlled operations on subsystems $A$ and $B$, with subsystem $C$ acting as the control register. Finally, another general unitary $U_C$ is applied to subsystem $C$.

\begin{figure}[h]
    \centering
    \includegraphics[width=0.5\linewidth]{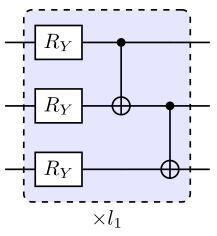}
    \caption{Illustration of the $V_i$ layer used in the construction of the unitary $V_C$ defined in Eq.~\eqref{eq:V_and_U}. Each $V_i$ consists of applying an $R_Y$ rotation gate to each qubit, followed by CNOT gates between neighboring qubits. The figure shows an example with 3 qubits, but this structure can be directly extended to $n$ qubits.}
    \label{fig:VC_illustration}
\end{figure}

\begin{figure}[h]
    \centering
    \includegraphics[width=0.8\linewidth]{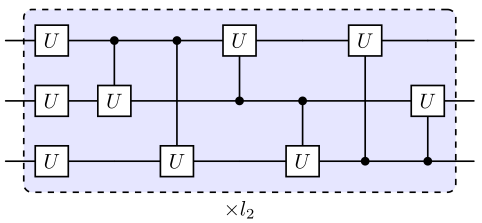}
    \caption{Layer $U_i$ used in the construction of the unitary $U_C$ defined in Eq.~\eqref{eq:V_and_U}. This layer begins with the application of a general rotation gate to each qubit, followed by the application of general controlled gates between all pairs of qubits.
}
    \label{fig:UC_illustration}
\end{figure}

An important aspect to highlight is that, as in other VQAs, the proper choice of unitaries is crucial for the performance of the algorithm. However, this choice poses a significant challenge due to the large number of possibilities. In this work, for simplicity, we define the unitaries $V_C$ and $U_C$ as follows:
\begin{equation}
    V_C = \prod_{i=1}^{l_1} V_i(\pmb{\theta}_i) \quad \text{and} \quad U_C = \prod_{i=1}^{l_2} U_i(\pmb{\phi}_i),\label{eq:V_and_U}
\end{equation}
where $V_i$ and $U_i$ are unitaries constructed from a set of quantum logic gates, parametrized respectively by the vectors $\pmb{\theta}_i$ and $\pmb{\phi}_i$. Figs.~\ref{fig:VC_illustration} and \ref{fig:UC_illustration} show the structure adopted for each $V_i$ and $U_i$, respectively.

Additionally, to demonstrate how the chosen parametrizations influence the algorithm's performance, we also consider, in some cases, the use of the \texttt{ArbitraryUnitary} function provided by the PennyLane library to construct $U_C$. While the definition of $U_C$ given in Eq.~\eqref{eq:V_and_U} is general, it is still constrained by the specific set of gates employed.  In contrast, the construction via \texttt{ArbitraryUnitary} yields the most general unitary possible for a given number of qubits.

It is important to note, however, that the practical applicability of the \texttt{ArbitraryUnitary} function is limited, as its implementation on real quantum hardware is hindered by the exponential growth in the number of gates required. Nonetheless, the results obtained with this function can serve as a theoretical benchmark, indicating the optimal performance that the algorithm could achieve as the unitary $U_C$, defined in Eq.~\eqref{eq:V_and_U}, approaches the more general form implemented via \texttt{ArbitraryUnitary}.

\section{Additional results}
\label{SM:Additional_results}

\begin{figure}[b]
    \centering
    \includegraphics[width=0.5\linewidth]{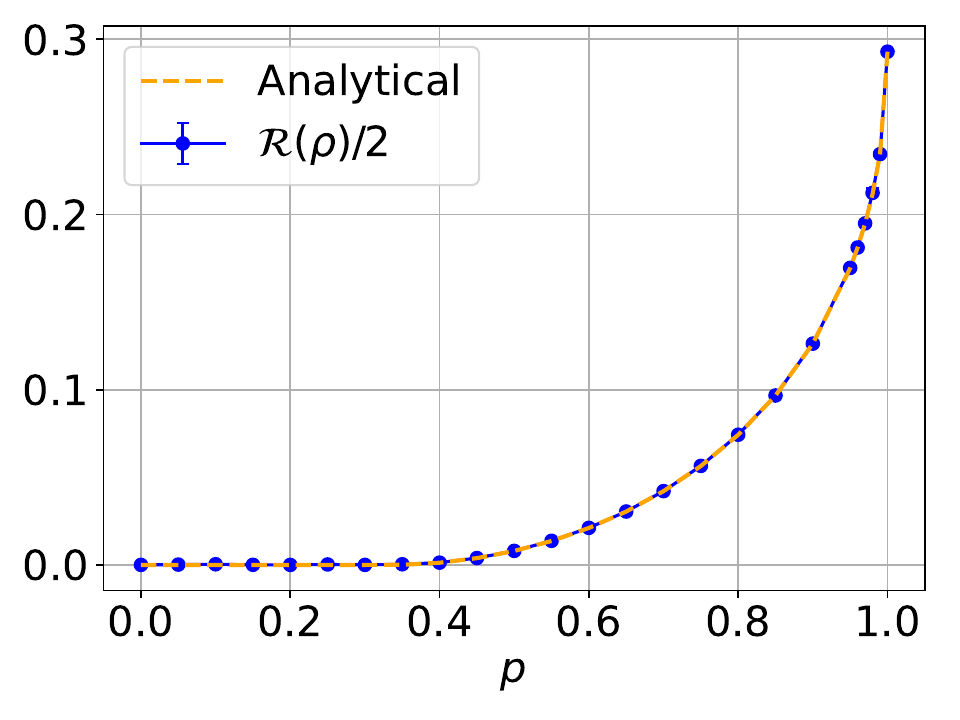}
    \includegraphics[width=0.5\linewidth]{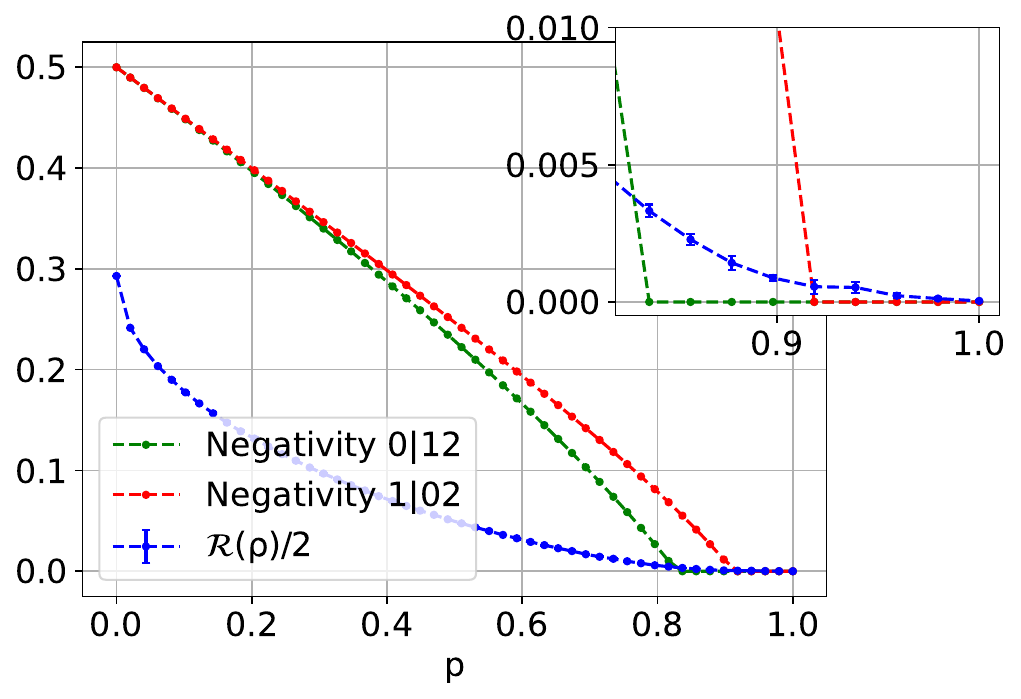}
    \includegraphics[width=0.43\linewidth]{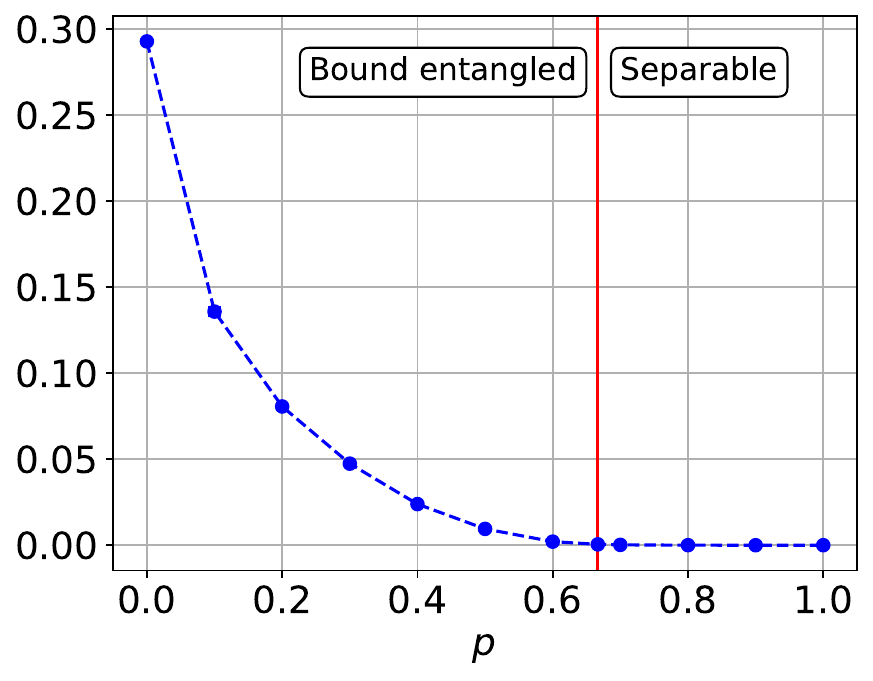}
    \caption{Top panel: Behavior of $ \mathcal{R}/2$ using the unitary $U_C$ generated by the \texttt{ArbitraryUnitary} function from Pennylane. In this case, we analyze the behavior of $ \mathcal{R}/2$ considering the state described in Eq. \eqref{eq:wenerState}. Left Panel: Behavior of $ \mathcal{R}/2$ using the unitary $U_C$ generated by the \texttt{ArbitraryUnitary} function from Pennylane. In this case, we analyze the behavior of $ \mathcal{R}/2$ considering the state described in Eq. \eqref{eq:lineargraphstate}. Right panel: Behavior of $ \mathcal{R}/2$ using the unitary $U_C$ generated by the \texttt{ArbitraryUnitary} function from Pennylane. In this case, we analyze the behavior of $ \mathcal{R}/2$ considering the state described in Eq.\eqref{eq:noiseSmolinState}.}
    \label{fig:graph_for_werner_state_using_ArbitraryUnitary}
\end{figure}

In this appendix, we present additional results that complement the main text. Specifically, the panels of Fig.~\ref{fig:graph_for_werner_state_using_ArbitraryUnitary}
show the entanglement estimation obtained by our algorithm. In these three cases, the unitary $U_C$ was constructed using the \texttt{ArbitraryUnitary} function.

For the Werner state, shown in the top panel of Fig.~\ref{fig:graph_for_werner_state_using_ArbitraryUnitary}, the results are consistent with those presented in the main text (see Fig.~\ref{fig:graph_for_werner_state}). For the three-qubit state, shown in the left panel of Fig.~\ref{fig:graph_for_werner_state_using_ArbitraryUnitary} and described by Eq.~\eqref{eq:lineargraphstate}, a slight difference can be observed, which can be attributed to the optimization process itself and does not indicate a limitation of our method. Finally, for the Smolin state -- defined in Eq.~\eqref{eq:noiseSmolinState} and illustrated in the right panel of Fig.~\ref{fig:graph_for_werner_state_using_ArbitraryUnitary} -- a more noticeable difference is observed, particularly for $p \ge 2/3$. In this regime, when using the unitary $U_C$ constructed via \texttt{ArbitraryUnitary}, the resulting entanglement is zero, in agreement with the theoretical results reported in Ref.~\cite{Smolin}. In contrast, when employing the $U_C$ defined in Eq.~\eqref{eq:V_and_U}, the entanglement values are slightly above zero, although still very close to it. These results clearly demonstrate the impact of the unitary $U_C$ choice on the algorithm’s performance--an expected outcome, now quantitatively confirmed.


A relevant aspect to be highlighted in this result concerns the problem of barren plateaus (BPs), which, as widely discussed in the literature, constitutes one of the main challenges faced by VQAs, significantly limiting their trainability and, consequently, their practical applicability. The results presented in Fig.~\ref{fig:graph_for_smolin_state} and in the right panel of Fig.~\ref{fig:graph_for_werner_state_using_ArbitraryUnitary} are particularly significant, as they show that the algorithm's performance is strongly associated with the expressiveness of the ansatz, which is expected, since the greater the expressiveness of the parametrization, the broader the set of states that can be generated and, therefore, the higher the chances of approaching the optimal result. However, as already established in the literature \cite{Holmes_2022,Friedrich_2023,Ragone}, this same degree of expressiveness is directly related to the occurrence of BPs, so that for the practical application of the proposed algorithm, it becomes essential to achieve an appropriate balance between expressiveness and trainability, ensuring that the ansatz is sufficiently expressive to capture the correct solution without compromising the feasibility of training.

\begin{figure}[t]
    \centering
    \includegraphics[width=1\linewidth]{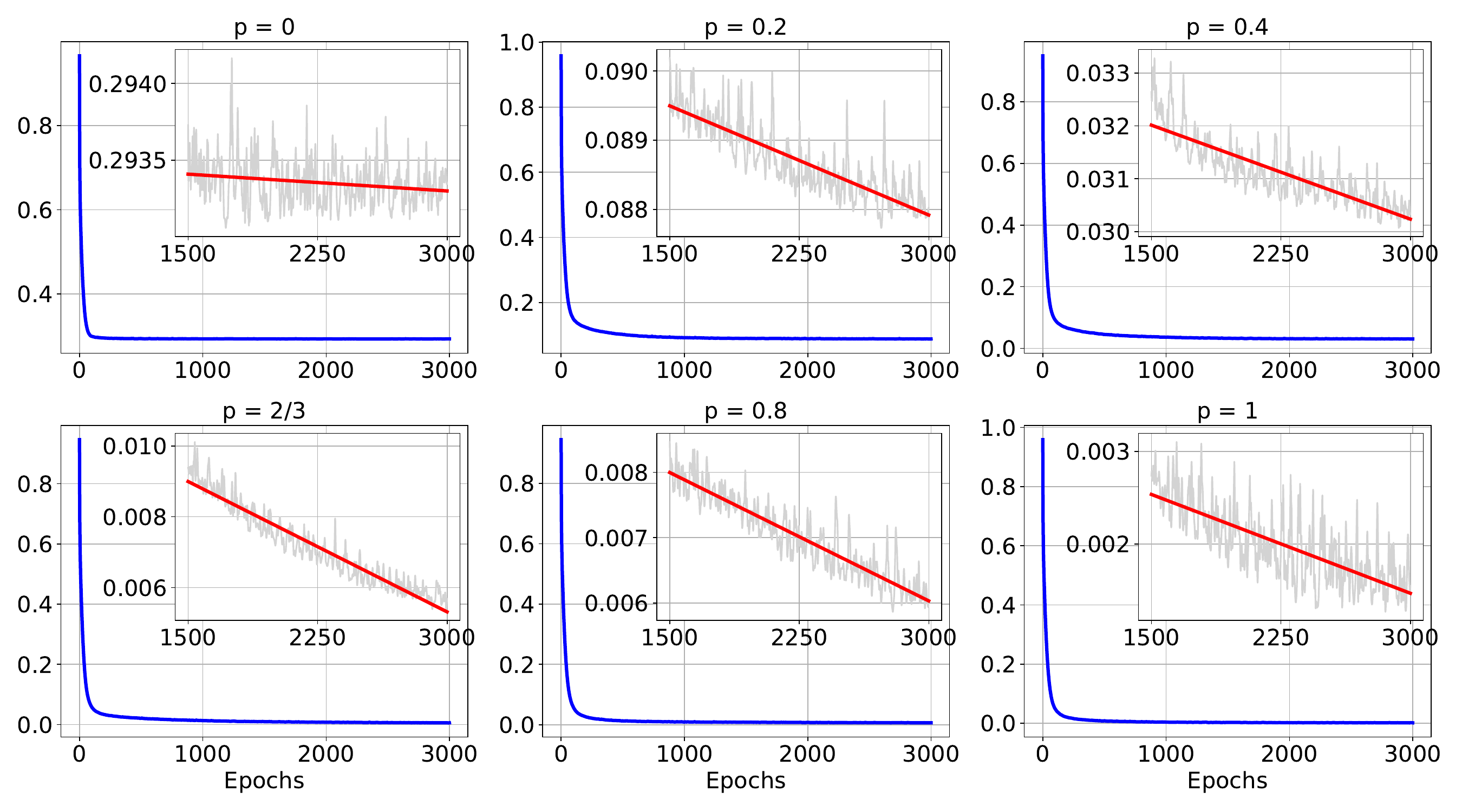}
    \caption{\footnotesize Behavior of the objective function $\mathcal{R}/2$ for the state in Eq.~\ref{eq:noiseSmolinState}, evaluated at different values of $p$ during training. The orange line marks the separability threshold. Each subplot includes a zoomed-in view (top right) highlighting the late-stage optimization trend in red.}
    \label{fig:grafico_p_variado_estado_de_smolin}
\end{figure}

\begin{figure}[t]
    \centering
    \includegraphics[width=1\linewidth]{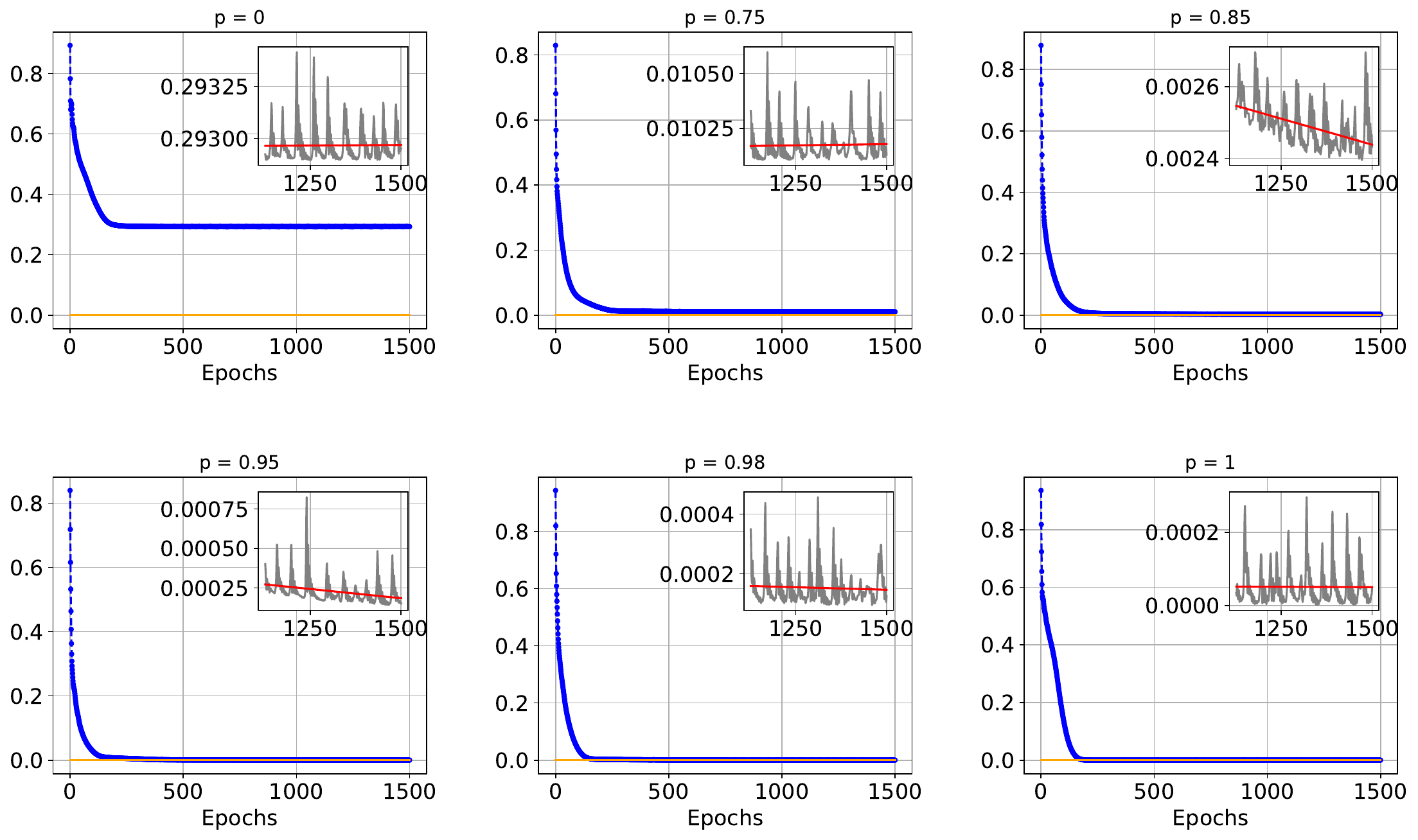}
    \caption{\footnotesize Objective function $\mathcal{R}/2$ during training for the state in Eq.~\ref{eq:lineargraphstate}, with different values of $p$. The orange line marks the separability threshold. Insets (top right) highlight, in red, the late-stage optimization trend.}
    \label{fig:grafico_p_variado_estado_grafo}
\end{figure}

Additionally, in Figs. \ref{fig:grafico_p_variado_estado_de_smolin} and \ref{fig:grafico_p_variado_estado_grafo}, we analyze the behavior of the function $ \mathcal{R}/2$ during training, considering the states described by Eqs. \eqref{eq:lineargraphstate} and \eqref{eq:noiseSmolinState} for different values of $p$. For these cases, the unitary $U_C$ used is the one illustrated in Fig. \ref{fig:UC_illustration}.

In Fig.~\ref{fig:grafico_p_variado_estado_de_smolin}, we present the behavior of $ \mathcal{R}/2$ for different values of $p$ in the case of the Smolin state, defined in Eq.~\eqref{eq:noiseSmolinState}. We observe that, in general, $ \mathcal{R}/2$ converges rapidly during the first 1000 epochs. However, after this point, the convergence slows significantly. This slowdown occurs because $ \mathcal{R}/2$ is already close to its ideal value, resulting in smaller gradients. Since the parameter updates depend on these gradients, the optimization process becomes slower. This effect is clearly visible in the zoomed-in plots, where the red lines show that, although the difference in $ \mathcal{R}/2$ between epochs $t$ and $t+1$ is small, there is still a consistent decreasing trend. Finally, Fig.~\ref{fig:grafico_p_variado_estado_grafo} shows the same analysis for the state defined in Eq.~\eqref{eq:lineargraphstate}.

\section{Comparison with other methods}
\label{sec:comparison}




In this appendix, with the aim of contextualizing the proposal of this work, we present a comparison among different approaches previously described in the literature. In particular, we focus our analysis on studies employing machine learning models, both classical and quantum, as well as on Variational Quantum Algorithm (VQA) methods, due to their conceptual similarities with the approach developed in this research.

The comparison is summarized in Table \ref{table:comparison}, where several key characteristics are examined. Specifically, we evaluate: (T1) whether the method depends on a training dataset, a feature commonly associated with machine learning approaches; (T2) whether the model’s performance is sensitive to the quality of such data; (T3) whether the method requires state tomography, a procedure known for its high computational cost; (T4) whether it is capable of detecting or quantifying bound entanglement; (T5) whether it can be applied to multipartite systems; (T6) whether it can handle mixed states; and (T7) whether it allows for explicit entanglement quantification.

In recent years, several methodologies have been proposed with the purpose of detecting and quantifying quantum entanglement. Among these, notable examples include classical machine learning approaches \cite{LU_2018,ASIF_2023,PAN_2024,Lin_2023_2,Urena_2024}, quantum machine learning approaches \cite{QIU_2019,Lin_2023}, and VQA-based algorithms \cite{Moller22,Zambrano,Philip}. Strategies based on machine learning generally involve training models on large datasets, often consisting of hundreds of thousands of samples. Consequently, the training process tends to present a high computational cost. In contrast, VQA-based approaches do not require pre-existing datasets, since training is performed directly on the states of interest.

An important aspect of machine learning-based approaches concerns the quality of the training dataset, which has a decisive influence on the model’s performance. For instance, when training a model designed to classify whether a state is separable or entangled, the algorithm learns patterns present in the provided data and uses them to infer the nature of new states. Thus, the quality and diversity of the training data determine the model’s generalization capability. If the dataset is restricted to a specific class of states, the model will tend to perform well only within that class. Similarly, if the training considers only one type of entanglement, the model will hardly be able to identify other forms of quantum correlation. Moreover, since many of these approaches employ the Positive Partial Transpose (PPT) criterion during the generation of training data, the inherent limitations of this criterion are often reproduced in the resulting models, which may compromise or even preclude the detection of bound entanglement.

Another challenge faced by machine learning-based approaches concerns the need for quantum state tomography (QST). Although it is often claimed that these methods do not require full QST, since they rely only on partial information about the state, in several cases \cite{ASIF_2023,PAN_2024,Urena_2024} QST remains indispensable, as the input data required by the models cannot be obtained otherwise. Furthermore, in other works \cite{LU_2018,Lin_2023_2,QIU_2019,Lin_2023}, the necessity of QST is not clearly addressed, since even when the trained model itself does not rely on complete QST, the generation of the states used to build the training dataset would still require it. A possible counterargument is that the training data are generated classically; however, this imposes severe practical limitations. Specifically, if the states used to create the training data are generated classically, their dimensionality will be limited by the capacity of classical computers to handle large dimensions. Consequently, even though these models may theoretically be applicable to multipartite systems, their practical use remains limited to low-dimensional systems \cite{ASIF_2023,PAN_2024,Lin_2023_2,QIU_2019}.

\begin{table}[h!]
\setlength{\tabcolsep}{2pt}
\begin{center}
\begin{tabular}{ | c |c | c | c | c | c | c | c |} 
  \hline
  \textbf{Method} & \textbf{T1}           & \textbf{T2} & \textbf{T3} & \textbf{T4} & \textbf{T5} & \textbf{T6} & \textbf{T7} \\
  \hline

  Ref. \cite{Philip} & \greencheck & \greencheck & \greencheck & Unclear & \greencheck & \greencheck & \redcross \\
  \hline

  Refs. \cite{Moller22, Zambrano} & \greencheck & \greencheck & \greencheck & \redcross & \greencheck & \redcross  & \greencheck \\
  \hline
  
  Ref. \cite{LU_2018} & \redcross & \redcross & Unclear & Unclear & \redcross & \greencheck & \redcross \\
  \hline
  
  Ref. \cite{ASIF_2023} & \redcross & \redcross & \redcross & \redcross & \greencheck -  BD & \greencheck & \redcross \\
  \hline
  
  Ref. \cite{PAN_2024} & \redcross & \redcross & \redcross & Unclear & \greencheck -  BD & \greencheck  & \greencheck \\
  \hline

  Ref. \cite{Lin_2023_2} & \redcross & \redcross & Unclear & Unclear & \greencheck -  BD & \greencheck  & \greencheck  \\
  \hline

  Ref. \cite{Urena_2024} & \redcross & \redcross & \redcross & \redcross & \redcross & \greencheck & \redcross \\
  \hline

  Ref. \cite{QIU_2019} &  \redcross &  \redcross & Unclear & Unclear & \greencheck -  BD & \greencheck & \redcross \\
  \hline

  Ref. \cite{Lin_2023} & \redcross & \redcross & Unclear & Unclear & \redcross & \greencheck & \greencheck \\
  \hline
  
  

  \textbf{EvQA} & \greencheck & \greencheck & \greencheck & \greencheck & \greencheck & \greencheck & \greencheck \\
  \hline
\end{tabular}
\caption{Criteria used for comparison: T1 - Does not depend on training data; T2 - Data quality does not influence results; T3 - Does not depend on quantum state tomography; T4 - Is able to quantify bound entanglement; T5 - Is applicable to multipartite systems; T6 - Is applicable to mixed states; T7 - Is capable not only of detecting but also of quantifying entanglement. Here, BD denotes "but may be difficult for large systems."}
\label{table:comparison}
\end{center}
\end{table}




\end{document}